\documentclass[final]{aipproc}
\usepackage{natbib}
\layoutstyle{6x9}

\def\gsim{\lower 2pt \hbox{$\, \buildrel {\scriptstyle >}\over
{\small \sim}\,$}}
\def\gtrsim{\lower 2pt \hbox{$\, \buildrel {\scriptstyle >}\over
{\small \sim}\,$}}
\def\lsim{\lower 2pt \hbox{$\, \buildrel {\scriptstyle <}\over
{\small \sim}\,$}}
\def\lesssim{\lower 2pt \hbox{$\, \buildrel {\scriptstyle <}\over
{\small \sim}\,$}}

\def\fuse{{\sl FUSE}}

\def\xmm{{\sl XMM-Newton}}

\def\suzaku{{\sl Suzaku}}
\def\chandra{{\sl Chandra}}
\def\Chandra{{\sl Chandra}}

\def\civ{C~{\small IV}}

\def\oi{O~{\small I}}
\def\oii{O~{\small II}}
\def\oiii{O~{\small III}}

\def\oviii{O~{\small VIII}}
\def\ovii{O~{\small VII}}
\def\ovi{O~{\small VI}}
\def\oi{O~{\small I}}
\def\oii{O~{\small II}}

\def\oiii{O~{\small III}}

\def\neix{Ne~{\small IX}}

\def\fexvii{Fe~{\small XVII}}

\begin{document}

\title{Global Hot Gas in and around the Galaxy}

\author{Q. Daniel Wang}
{address={Department of Astronomy, University of Massachusetts, Amherst}}

\begin{abstract}

The hot interstellar medium traces the stellar feedback and its
role in regulating the eco-system of the Galaxy.
I review recent progress in understanding the medium, based largely on 
X-ray absorption line spectroscopy, complemented by 
X-ray emission and far-UV \ovi\ absorption measurements.
These observations enable us for the first time to characterize the global spatial, 
thermal, chemical, and kinematic properties of the medium. 
The results are generally consistent with what have been inferred from 
X-ray imaging of nearby galaxies similar to the Galaxy.
It is clear that diffuse soft X-ray emitting/absorbing gas with a 
characteristic temperature of $\sim 10^6$ K resides primarily
in and around the Galactic disk and bulge. In the
solar neighborhood, for example, this gas has a characteristic vertical scale height
of $\sim 1$ kpc. This conclusion does 
not exclude the presence of a larger-scale, probably much hotter,
and lower density circum-Galactic hot medium, 
which is required to explain observations of various high-velocity clouds. 
This hot medium may be a natural product of the stellar 
feedback in the context of the galaxy formation and evolution.
\end{abstract}

\maketitle

\section{Introduction}

An understanding of the Local Bubble (LB) must be in the context of
the global interstellar medium (ISM). In particular, the very evidence for
the presence of $\sim 10^6$ K gas in the LB is now questioned, because of 
the uncertainty in the contribution of the solar wind charge exchange (SWCX) 
to the observed diffuse soft X-ray background (e.g., see the contribution 
by Koutroumpa et al.; \cite{shelton2007}). In the mean time, breakthrough observational progress has 
been made recently in characterizing the global hot ISM --- a topic that I am 
reviewing here. I will further compare this characterization with the global perspectives 
obtained from observing hot gas in and around 
nearby galaxies. Finally, I will discuss the implications of the results,  
exploring the role of the global hot gas in regulating the Galactic ecosystem
and the evolution of the Galaxy in general.

\section{Observations}

\subsection{X-ray Emission}
Until recently, our knowledge about the 
hot ISM ($T \gsim$ a few $\times 10^5$ K)  came almost exclusively from 
various broad-band observations of the diffuse soft X-ray background.
Maps generated from the ROSAT all-sky survey 
(RASS; \cite{1997ApJ...485..125S}), for example, show the angular
distribution of the  background intensity in various energy bands.
At energies $\lesssim 0.3$ keV, much of the background, if not due to the SWCX, 
should be produced by hot gas inside the LB. But as demonstrated in several 
measurements of X-ray shadows cast by cool gas
clouds at known distances, a significant contribution of the
background at high Galactic latitudes can have a more distant origin 
\citep[e.g.,][]{2000ApJ...543..195K}.
In addition to an extragalactic component, consisting of mostly faint AGNs,  
a contribution is also expected from a thick hot gaseous Galactic disk and/or a 
large-scale circum-Galactic hot medium on scales $\gtrsim 10$ kpc. The distant contribution 
must dominate at higher energies. In particular, the X-ray 
background in the 0.5-2 keV range shows a general enhancement towards the inner 
part of the Galaxy. The nature of this enhancement 
is still greatly uncertain, but probably represents a combination
of a nearby superbubble and an outflow from the Galactic nuclear region. 
Away from this inner region, the 0.5-2 keV background appears rather uniform,
even across the Galactic plane. Such a uniformity provides the first 
indication for a Galactic disk contribution that more-or-less compensates the
X-ray absorption by cool gas. This disk contribution most likely represents
an accumulated contribution from stellar emission 
(primarily cataclysmic variables and coronally active binaries; 
\citep[and references therein]{2008A&A...483..425R}
as well as the hot ISM.

Spectroscopic observations have provided additional 
information about the nature of the soft X-ray background emission. For example,
rocket experiments with instruments of sufficient spectral 
resolution have been used to detect individual lines or line complexes 
\citep[e.g.,][]{2002ApJ...576..188M}. At high Galactic latitudes, about 50\% of the background 
in the RASS $\sim 0.75$ keV band is shown to be thermal in origin, while the rest can be 
accounted for by the nonthermal extragalactic AGN contribution. 
The spectra from such experiments were, however, accumulated from large 
swaths of the sky. More recently, 
reasonably good spectroscopic imaging data can be obtained from the X-ray 
CCDs on-board \suzaku\ with a much improved spectral resolution, compared with
those on \chandra\ and \xmm. Fig.~1c presents a {\sl Suzaku} XIS diffuse X-ray spectrum obtained
in a field adjacent to the sight line of LMC X-3 \cite{2008arXiv0808.3985Y} 
--- an X-ray binary
for which both UV and X-ray absorption line spectroscopic data are available 
for comparison \cite{2005ApJ...635..386W}.
The emission spectrum clearly shows \ovii, \oviii, and \neix\ K$\alpha$ lines, 
suggesting the presence of an optically thin thermal plasma with an average temperature 
of $\sim 2.4 \times 10^6$ K.

The X-ray emission data alone, however, give little information on the 
global hot ISM. In particular, no kinematic and distance information 
can be extracted. The interpretation of the data very much depends on 
the assumptions of the density and temperature uniformity as well as of 
the relative cool (X-ray-absorbing) and hot gas distribution. 

\subsection{X-ray Line Absorption}

Thanks to \chandra\ and \xmm, we now have a new 
powerful tool --- X-ray absorption line spectroscopy --- 
to study the global hot gas. Unlike X-ray emission, which is sensitive 
to the emission measure of the hot gas, absorption
lines produced by ions such as \ovii, \oviii, and \neix\ 
directly probe their column densities, 
which are proportional to the mass of the gas and are
sensitive to its thermal, chemical and kinematic properties. One also does not need to 
deal with the SWCX, because it contributes little to the line absorption. 
Furthermore, the line absorption, independent of photo-electric absorption by 
cool gas, samples hot gas unbiasedly along a sight line. Therefore, the 
absorption line spectroscopy, particularly in combination with the analysis of the emission
data, now enables us to explore for the first time the global hot ISM in the Galaxy. 

X-ray absorption lines produced by diffuse hot gas were first detected
in the spectra of several bright AGNs (PKS~2155--304, 3C~273, and Mrk~421). 
While the very detection of such lines at non-zero redshifts remains to be controversial, 
the presence of the absorption by hot gas at $z \sim 0$ is firm 
\citep[e.g.,][and references therein]{2003ASSL..281..109R,2005ApJ...631..856W,2005ApJ...624..751Y}
The spectral resolution of the grating observations,
however, is still limited to about $\sim 500 {\rm~km~s^{-1}}$, although
 with a sufficient counting statistics the centroid of a line
can be determined to a much higher accuracy. Therefore, kinetically, the
absorbing hot gas can be anywhere within a few Mpc around the Galaxy.
But physically, it has been argued forcefully that the effective spatial scale 
of the gas cannot be on the Local Group scale of $\sim 1$ Mpc; otherwise, the amount of the baryon matter in the hot gas
would exceed the expected baryon mass in the enclosed region 
\citep[e.g.,][]{2006ApJ...644..174F}. Furthermore, the hypothesis that
the gas may represent the intra-group gas in the Local Group of
galaxies may be ruled out because of the expected enhancement
of the absorption in the M31 direction is not observed \cite{2007ApJ...669..990B}.

A more direct constraint on the location of the $z \sim 0$ hot gas
comes from the absorption line spectroscopy of the sight line toward LMC X-3 
\citep[Fig.~1;][]{2005ApJ...635..386W}.  The detected \ovii\ and \neix\ K$\alpha$ 
absorption lines have equivalent widths similar to those
seen in the spectra of the AGNs. Most importantly, the absorption must occur 
within the source distance of 50 kpc and the line centroid is also inconsistent
with the systemic velocity of the Large Magellanic Cloud. 
Indeed, similar interstellar absorption lines have been 
detected in the spectra of Galactic X-ray binaries 
\citep[e.g.,][]{2004ApJ...605..793F, 2005ApJ...624..751Y}.
It is now clearly that the hot ISM is present {\sl globally} in the Galaxy and is 
the primary contributor to the observed $z \sim 0$ highly-ionized X-ray absorbers.

\begin{figure}[]
\includegraphics[height=9.8cm]{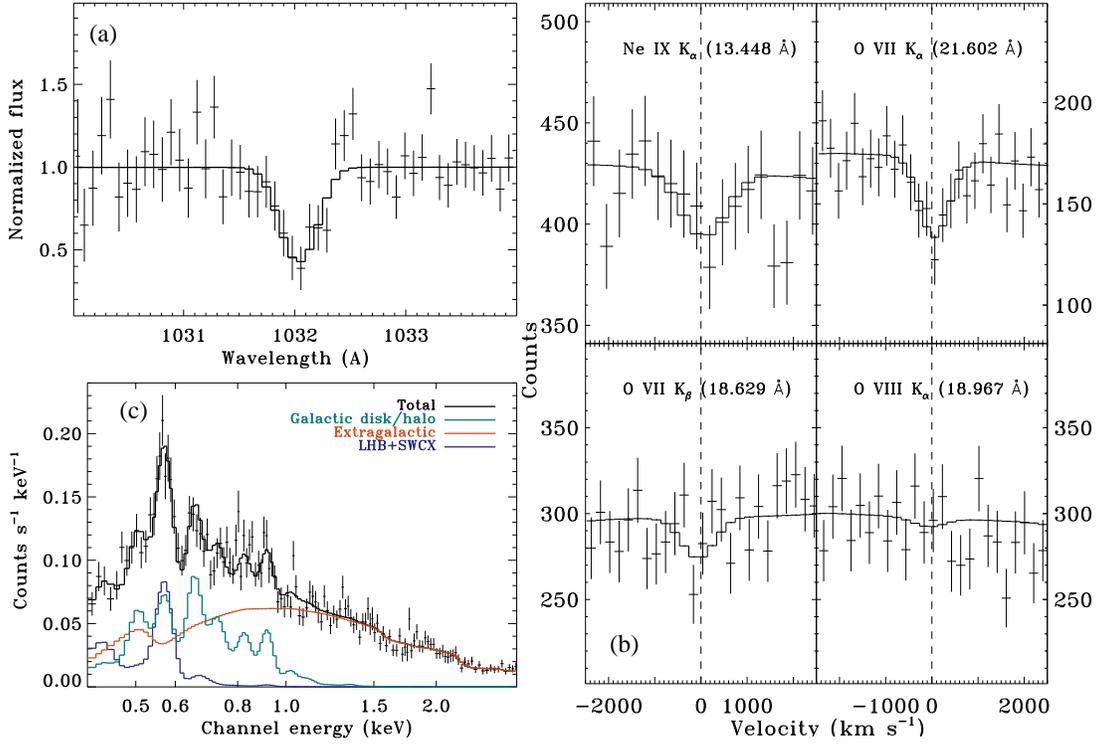}
  \caption{\small Spectroscopic data for the LMC X-3 sight line: (a)
 {\sl FUSE} detection of the \ovi\ absorption line, (b) the Neon and Oxygen
 absorption lines, as observed by {\sl Chandra}, and (c) the local X-ray
background spectrum from off-field {\sl Suzaku} observations.
The solid histograms in the plots represent a joint-fit of our 
non-isothermal model of the coronal gas to the data sets; panel
(c) includes separately the contributions from the extragalactic
background and the estimated charge exchange contribution, 
in addition to the hot ISM.
    \label{fig:exist}
  }
\end{figure}

\subsection{Far-UV Line Absorption and Emission}

While the X-ray emission and absorption trace hot gas, 
far-UV spectroscopy is sensitive to gas at a few times $10^5$ K. 
This so-called intermediate-temperature gas
is particularly important for understanding the interplay between cool and
hot gases. Far-UV spectroscopy also typically has spectral 
resolution higher enough to resolve individual emission/absorption lines, providing
valuable kinematic information. 
So far most of such far-UV spectroscopy has been done in absorption for lines
such as \ovi\ and \civ\ \citep[e.g.,][]{2003ApJS..146..125S}.
The bulk of such absorption is known to be in a Galactic 
disk with a vertical scale height of a couple of kpc.
While the population of \civ\ 
could be partly due to photon-ionization, \ovi, 
sensitively probed with {\sl FUSE} as the
1031.9  and 1037.6 \AA\ resonance line doublet, should originate 
predominantly in collisionally ionized plasma.
The doublet has also been 
detected in diffuse emission, typically based on very long exposures with
\fuse\ because of its small aperture ($30^{\prime\prime} \times
30^{\prime\prime}$). Unlike the absorption, however, the 
emission intensity is subject to extinction, the correction 
of which can be rather uncertain. Also few emission observations were taken in close vicinities of 
the absorption sight lines \citep[e.g.,][]{2006ApJ...647..328D}. Therefore, the comparison between 
the absorption and emission data has been difficult. Furthermore, the 
inference of the relevant physical parameters from the far-UV data alone 
requires an assumption of the gas temperature, often taken to be $\sim 3 \times 10^5$~K, at which
the  \ovi\ ionic fraction peaks, assuming a collisional ionization equilibrium. Under this 
assumption, one neglects the \ovi\ contribution from thermally more stable and
thus more abundant gas at higher temperatures.

\section{Characterization of the Global Hot Gas}

The rich data sets described above provide an excellent opportunity
to characterize the global hot gas. An analysis tool has been developed to explore the 
data sets, especially the new X-ray absorption line spectroscopic capability  
\cite{2005ApJ...624..751Y, 2008arXiv0808.3985Y}. This tool allows
us to joint fit multiple absorption lines (detected or not) as well as the emission 
measurements, spectroscopic or broad-band. 
Such joint fit maximizes the constraints from available data
on spatial, thermal, chemical, and/or kinematic parameters of the hot gas with
proper error propagation. Our analysis has so far assumed a collisional ionization equilibrium
which is generally suitable for the global hot ISM.
Here I highlight key results from such analysis for hot gas in the 
Galactic disk (defined to be within a few kpc of the 
Galactic plane), the Galactic bulge (a few kpc from the Galactic center), and the
circum-Galactic medium (on scales beyond the Galactic disk and bulge, up to 
$\sim 200$ kpc or the virial radius of the Galaxy).

\subsection{Hot Gaseous Disk}

The sight-line/field of LMC X--3 (Galactic coordinates $l, b =279^\circ, -46^\circ; D=50$ kpc)
provides an excellent sample of the  hot gas away from the Galactic inner region. 
This bright black hole X-ray binary
is located far outside the main body of the LMC \citep[see Fig.~1 of][]{2005ApJ...635..386W}. 
The systemic velocity of the source ($+310
{\rm~km~s^{-1}}$) further makes it relatively straightforward to
distinguish any potential absorption that is local to the binary or to
the LMC. The analysis of the comprehensive far-UV/X-ray data available 
for this sight-line/field (Fig.~1)
has given the following key results \cite{2005ApJ...635..386W,2008arXiv0808.3985Y}:
(1) Both far-UV \ovi\ and X-ray \ovii\
absorption lines show little offset from the local standard of rest, consistent with a
Galactic interstellar origin; (2) 
Both X-ray emission and absorption data can be fitted well with a hot
gaseous disk model with the temperature and density decreasing
exponentially with the vertical distance from the Galactic plane,
i.e., $n=n_0e^{-z/h_n}$ and $T=T_0e^{-z/h_T}$, where $n_0\sim1.4\times10^{-3}~{\rm cm^{-3}}$,
 $h_n\sim2.8$ kpc, $T_0\sim3.6\times10^6$ K, and $h_T\sim1.4$ kpc; (3) This
X-ray-data-constrained model can naturally account for all the
observed far-UV \ovi\ absorption.
These results agree well with those obtained for the Mkr 421 sight line, 
which is also away from the Galactic inner region and is analyzed in a 
similar fashion \cite{2007ApJ...658.1088Y}. 

Furthermore, a statistical analysis of the hot gas column densities along the sight
lines toward a sample of Galactic and extragalactic sight lines gives a similar disk scale 
height for the hot gas. This scale height is also consistent with those inferred from 
similar statistical analyses of far-UV absorption lines and pulsar dispersion measures
\cite{2005ApJ...624..751Y}. 

The sight line toward the low-mass X-ray binary (LMXB) 4U 1820-303 ($l, b =3^\circ, -8^\circ; D=8$ kpc), for example, 
presents a relatively low Galactic latitude view of the global ISM through the entire inner Galactic disk 
\cite{2004ApJ...605..793F,2006ApJ...641..930Y}. \Chandra\ grating observations show 
the presence of \fexvii\ L transition as well as \oi, \oii, \oiii, \ovii, \oviii, 
and \neix K$\alpha$  
 absorption lines, which are apparently interstellar in origin. The LMXB is 
super-compact; the large ionization parameter in the immediate vicinity of 
the binary greatly reduces the chance for a significant local contribution to 
the observed absorption lines, consistent with their constancy, moderate 
ionization state, and insignificant width/shift. 
Furthermore, the LMXB resides in the globular cluster 
NGC 6624, which also contains two radio pulsars with the dispersion 
measures giving a total free electron column density useful 
for an absolute abundance measurement of ionized gas.
Based on these data, \cite{2006ApJ...641..930Y} obtain the column densities of the neutral, 
warm ionized, and hot phases of the ISM along the sight line. 
They find that the mean oxygen abundance in the neutral atomic phase
is significantly lower than in the ionized phases. 
This oxygen abundance difference is apparently a result of 
molecule/dust grain destruction and recent metal enrichment in the warm ionized and hot phases. 

Hot gas properties toward the Galactic inner regions seem to be significantly different from 
those in the solar neighborhood. Both the measured mean temperature 
and velocity dispersion are substantially higher along the sight lines
toward LMXBs in the inner regions: $\sim 10^{6.3}$ K and $200 
{\rm~km~s^{-1}}$ vs. $\sim 10^{6.1}$ K and $50 {\rm~km~s^{-1}}$, 
respectively \cite{2005ApJ...624..751Y,2007ApJ...666..242Y}. 
While the X-ray absorption line profiles are not resolved
with the existing grating observations, the velocity dispersion is 
inferred from the \ovii\ K$\alpha$ and K$\beta$
ratio --- a measure of the relative saturation that is sensitive to the line
broadening. The substantially large velocity dispersion toward the inner regions indicates strong
turbulent and/or large differential bulk motion of hot gas in and around
the Galactic bulge.

\subsection{Hot Gaseous Bulge}

To probe the location dependence of the global hot ISM in the Galaxy,
One can study the differential X-ray absorption/emission properties along 
multiple sight lines. \citet{2007ApJ...666..242Y} have reported
such a study of the soft X-ray background enhancement toward the inner region 
of the Galaxy. 3C 273 and Mrk 421 are two AGNs with their sight lines on and off 
the enhancement, but at similar Galactic latitudes. 
The diffuse 3/4 keV emission intensity is about 3 times higher toward 3C 273 
than toward Mrk 421. \chandra\ grating observations of these two AGNs are used
to detect X-ray absorption lines 
(e.g., \ovii\  K$\alpha$, K$\beta$, and \oviii\ K$\alpha$ transitions 
at $z \sim 0$). The mean hot gas thermal and kinematic properties 
along the two sight lines are significantly different. By subtracting the combined foreground 
and background contribution, as determined along the Mrk 421 sight line, they 
isolate the net X-ray absorption 
and emission produced by the hot gas associated with the enhancement in the direction of 3C 273. From a joint analysis of these differential data sets, they obtain the temperature, dispersion velocity, and hydrogen column density as 2.0 $\times 10^6$ K, 
$216 {\rm~km~s^{-1}}$, and 2.2 $\times 10^{19} {\rm~cm^{-2}}$, which are
significantly different from the measurements for hot gas away from
the enhancement. The effective line-of-sight extent of the gas is also 
constrained to be in the range of $1-10$ kpc, strongly suggesting that the 
enhancement along the 3C 273 sight line is primarily due to 
a Galactic central phenomenon. 

The difference between the Galactic disk and bulge regions also appears 
in the metal abundance of the hot gas. Irons
are known to be strongly depleted in the local hot gas, probably by a factor of
$\gtrsim 10$, based on X-ray emission line spectroscopy 
\cite{2002ApJ...576..188M,2005ApJ...623..911H}. This depletion is also consistent with the 
lack of a significant \fexvii\ L transition line detection away from the Galactic central
region (Y. Yao, private communications). Presumably the depleted irons reside
in certain dust grains (or their sturdy cores), which have survived the thermal 
sputtering. The presence of such grains can be a very important 
energy sink of hot gas. However, along the sight line of 4U 1820-303, 
the estimated Fe/Ne abundance ratio is about solar. This ratio can be
understood as the averaging of an extreme iron depletion in the disk
and a large enhancement in the bulge. The iron enrichment in the bulge  is expected 
because of frequent Type Ia SNe. 

\subsection{Circum-Galactic Hot Medium}

One can further use the depth differences of multiple sight lines to constrain the 
hot gas column density of the large-scale Galactic halo. \citet{2008arXiv0808.3985Y} 
have compared the detections of \ovii\ and \neix\ K$\alpha$ absorption lines along the sight 
line of 4U 1957+11 ($l, b =51^\circ, -9^\circ; D=10-25$ kpc), a persistent 
Galactic low-mass X-ray binary, and those toward extragalactic sight lines. They 
find that all the line absorptions can be attributed to the hot gas in a thick 
Galactic disk, accounting for the Galactic latitude-dependence. The \ovii\ 
column density of the circum-Galactic hot medium is constrained to be $N_{\ovii} 
< 5 \times 10^{15} {\rm~cm^{-2}}$ (95\% confidence limit). A tighter constraint with 
$N_{\ovii} 
< 2 \times 10^{15} {\rm~cm^{-2}}$ has been recently obtained with the comparison of the
Cyg X-2 and the Mrk 421 sight lines (Yao et al. 2009 in
preparation/private communication).

So far the most compelling, though indirect, evidence for the circum-Galactic hot 
medium comes from 
observations of high-velocity clouds \citep[HVCs; e.g.,][]{2003ApJS..146..165S}. 
HVCs in general represent a heterogeneous population,
resulting from Galactic disk fountains, gas stripped from
satellite galaxies, condensation from the hot medium, and 
possibly intergalactic clouds. A large sub-population of the HVCs
(e.g., those detected in 
absorption lines produced by highly ionized ions such as \ovi)
appears to be located far away from the Galactic disk ($D \gtrsim 10$ kpc), 
although tight distance constraints are few.
The velocity distribution of this sub-population indeed favors the circum-Galactic 
origin \cite{2005ApJ...623..196C}.
The HVCs also show a range of metallicity with the mean 
of $\sim 0.1$ solar, which is often used as a canonical value, although
the number of good measurements is still very limited. The hot medium is required to 
provide the pressure confinement of the HVCs; otherwise they would be dispersed quickly.
Also some of the HVCs, including those as part of the Magellanic Stream, 
show comet-like shapes, indicating stippling by the medium.
The considerations of the survivability of the Magellanic Stream puts a limit on
the density of the hot medium at $D \sim 50$ kpc from $\lsim 10^{-4} {\rm~cm^{-3}}$ based on dynamical arguments 
to $\lsim 10^{-5} {\rm~cm^{-3}}$ based on heating and evaporation arguments \citep[and references therein]{2000ApJ...529L..81M}. The temperature of the hot gas is not constrained, observationally. But a
temperature lower than a couple of $10^6$ K would lead to a rapid cooling of the medium,
for which no evidence is found in X-ray, and would be difficult
to provide the pressure confinement.
The presence of the hot medium is also necessary to produce the observed amounts of \ovi,
which resides mostly in a transitional phase 
between thermally more stable cool and hot phases and is typically
found at conductive interfaces, turbulent mixing layers, and/or cooling
flows. 

In short, the current interpretation of the HVCs 
demands the presence of the circum-Galactic hot medium with a very low density. 
The high temperature and low density also naturally explain the lack of detectable X-ray
emission and absorption from the medium and around other nearby galaxies similar 
to our own one.

\section{Discussion}

\subsection{Comparison with Hot Gas in Nearby Galaxies}

To fully comprehend the global hot gas in and around the Galaxy, we need to 
complement our insider's view, presented above, with external perspectives 
from observing nearby galaxies of similar properties.
To complement the study of the Galactic bulge, which is
largely obscured in the wavelength range from optical to soft X-ray, 
one can examine hot gas in nearby galactic bulges. Based on a careful analysis
of X-ray data on M31, \citet{2007ApJ...668L..39L}
demonstrated that the properties of hot gas 
in a galactic bulge can be  characterized if the contribution from X-ray binaries
can be excised, which typically have individual 
luminosities $\gsim 10^{36} {\rm~ergs~s^{-1}}$; the collective X-ray emissivity 
of fainter sources can be tightly constrained, because they 
should spatially follow the stellar (K-band) light intensity \cite{2008A&A...483..425R}. 

The unambiguous detection of the diffuse hot gas in 
and around the M31 bulge helps to understand the soft X-ray enhancement observed toward the inner
region of our Galaxy \cite{2007ApJ...668L..39L}. The temperature of the hot gas associated with the 
M31 bulge is similar to that with the Galactic bulge, as estimated
from the RASS in regions with $|b| \gtrsim 10^\circ$  
\cite{1997ApJ...485..125S}. But the X-ray emission of the Galactic bulge
appears to be more extended and about a few times more luminous than that of
the M31 bulge, presumably due to the feedback from recent extensive massive star formation
at the Galactic center \citep[e.g.,][and references therein]{2006JPhCS..54..115W}.  
Within $|b| \lesssim 10^\circ$,
the interstellar absorption is severe, little can be inferred reliably about the properties
of the hot gas. It is in this corresponding region in the M31 bulge 
that the diffuse soft X-ray intensity shows the steepest increase 
(by about one order of magnitude) toward the galactic center. 
Such a mid-plane enhancement of 
diffuse soft X-ray emission may also be present intrinsically in our 
Galactic bulge. The enhancement may be a result of the strong concentration
of the SN energy input and the mass-loading from cool gas.

\citet{2007EAS....24...59W} has reviewed recent work on the global hot gas around nearby normal 
galaxies, particularly edge-on ones. The results are
generally consistent with the above characterization for the hot gas
in our Galaxy. Observed diffuse X-ray emitting/absorbing gas typically does
not extend significantly more than 10 kpc away from galactic 
disks/bulges, except in nuclear starburst or very massive galaxies. 
The morphology of the diffuse X-ray emission as well as its correlation with the 
star formation rate and the stellar mass clearly show that the detected extraplanar
hot gas is primarily heated by the stellar feedback. In particular, recent
analysis indicates that the hot gas may have multiple components of significantly
different temperatures \cite{2003ApJ...598..969W,2008arXiv0807.3587L}. Outflows from 
current and recent star forming regions (with stellar ages $\lesssim 10^7$ yrs) 
could be strongly mass-loaded, producing a lower temperature component, while those from
older star forming regions in a galactic disk and from a galactic bulge may be 
responsible for the higher temperature component. But this scenario needs to be
tested further with a systematic study of the relationship between the star formation and
diffuse hot gas. Energetically, much of the expected stellar feedback seems to be
``missing'', at least not detected in X-ray emission 
\cite{2006MNRAS.371..147L,2007MNRAS.376..960L,2007ApJ...668L..39L}.
The observed X-ray luminosity of diffuse hot gas is typically less than a few \% of the energy
input from SNe. This missing feedback 
problem is particularly acute in so-called low $L_X/L_B$ 
bulge-dominated galaxies (typically Sa spirals, S0, and low mass ellipticals),
in which little cool gas can hardly hide or radiate the 
energy in other wavelength bands. Most likely, the missing 
energy is gone with galactic outflows. 

\subsection{The Role of Global Hot Gas in Shaping the Circum-Galactic Medium}

The presence of the global hot gas and its outflow can strongly
affect its gaseous structure and evolution of a galaxy. 
According to the current theory of the structure formation, galaxies similar to the 
Milky Way were formed and are still evolving from the mist of the 
intergalactic medium (IGM). However, the actual formation and evolution  process inside
individual galaxies remains largely uncertain, critically depending on the treatment of the feedback 
from stars and AGNs. Therefore, it is important to understand the global hot gas
as a tracer of the feedback in the context of galaxy formation and
evolution. 

One attempt in this direction is to understand the low density of 
the circum-Galactic hot medium. \citet{2008arXiv0803.4215T} 
demonstrate that the feedback from the stellar 
bulge can play an essential role in shaping the global hot gas. They have conducted 
1-D hydrodynamical simulations of the hot gas evolution based on a feedback model consisting
of two distinct phases: 1) an early starburst during the bulge formation and 2) 
a subsequent long-lasting mass and energy injection from stellar winds of low-mass stars and 
Type Ia SNe. An energetic outward blastwave is initiated by the starburst and is maintained and enhanced by the long-lasting stellar feedback. This blastwave can
heat up the surrounding medium to a scale much beyond the virial radius of the halo, thus the hot gaseous accretion can be completely stopped. In addition, the long-lasting feedback in the later phase powers a Galactic 
bulge wind that is reverse-shocked at a large radius and maintains the
circum-galactic hot medium. As the mass and energy injection decreases with time, 
the feedback may evolve to a subsonic and quasi-stable outflow, which is enough to prevent 
the development of a massive cooling flow. 
The two phases of the feedback thus re-enforce each-other's impact 
on the gas dynamics. The simulation results demonstrate that the stellar bulge feedback 
may provide a plausible solution to the missing stellar feedback problem and the over-cooling 
problem (over-predicting the amounts of cooled and cooling gas in and around a Milky Way-like  galaxy). 

To directly confront with the far-UV and X-ray emission/absorption measurements, one still needs to
conduct more detailed modeling of the hot gas in and around the bulges and disks of our Galaxy, M31,
and other similar galaxies. High spatial resolution 3-D simulations are  necessary to
account for density and temperature structures, which the emission is particularly sensitive to.
Preliminary simulations have shown that the mass-loading from cool
gas may play an important role in determining hot gas properties. Indeed,
observational evidence is present for this process: a detailed analysis of the \chandra\ ACIS data shows 
enhanced X-ray emission around the nuclear cool gas spiral of the M31 bulge, while
the detection of a strong OVI line absorption directly indicates the presence of
large amounts of
intermediate-temperature gas \cite{lww2008}. The magnetic field may also
be important in shaping the outflow of hot gas, as indicated by the bi-polar morphology of
the diffuse X-ray emission in and around the bulge of M31 \cite{2007ApJ...668L..39L}. 
These studies have
demonstrated that detailed modeling of the X-ray and far-UV data can be very useful 
to the understanding of how the feedback works in and around galaxies.

\begin{theacknowledgments}
I thank the organizers of the workshop for the invitation to give this
talk and am grateful to my students and collaborators (particularly Yangsen Yao, 
Zhiyuan Li, and 
Shikui Tang)  for their contributions to the work as reviewed above, which is 
partly supported by NASA/CXC under grants NNX06AB99G, NNG05GC69G, and GO5-6078X.
\end{theacknowledgments}

\bibliographystyle{aipproc}
\bibliography{bib}

\end{document}